\documentclass[a4paper,11pt]{article}
\usepackage{amsmath}
\usepackage[usenames]{color}
\usepackage[dvips]{graphicx}

\newcommand{\be}{\begin{equation}}
\newcommand{\ee}{\end{equation}}
\newcommand{\bea}{\begin{eqnarray}}
\newcommand{\eea}{\end{eqnarray}}
\newcommand{\nl}{\nonumber \\}
\newcommand{\bra}{\langle}
\newcommand{\ket}{\rangle}

\newcommand{\s}{\displaystyle{\not}}

\begin{document}

\begin{center}
\begin{Large}
{\bf Numerical Implementation of Generalized Unitarity}
\end{Large}
\end{center}

\vspace*{\baselineskip}

\begin{center}
{\bf  Petros Draggiotis\footnote{{\tt pdrangiotis@ugr.es \;  University of Granada after 1 October 2007 }} }  \\
{ \small University of Athens,Physics Department, Nuclear \& Particle Physics Section, Athens,Greece}
\end{center}

\vspace*{0.3cm}

\begin{abstract}
We present a numerical implementation of generalized unitarity. We will show that we are able to obtain
the box coefficients of any 1 loop gluonic amplitudes for an arbitrary helicity configuration and for any
number of external particles.
\end{abstract}

\section{Introduction}

One loop amplitudes is an important and very complicated part of a scattering cross section
calculation in high energy physics. Even for a small number of external particles the number of Feynman diagrams
grows really fast, when one adds just one more leg. During the last years, there has been a lot of progress in
understanding the complex structure of 1 loop amplitudes and new methods have helped us to work around
Feynman diagrams in getting to the result. These methods depend on the complex plane structure of an amplitude
when we continue all momenta to become complex.

The particular method that we focus, {\it generalized unitarity}, is a clever method of obtaining the coefficients
of the master integrals, that constitute the diverging part of an amplitude. Based on cutting the propagators in the
loop and obtaining a particular coefficient as product of tree amplitudes, the method lends itself naturally for an algorithmic
implementation. We have developed just such an implementation, which (as a first step) computes the box coefficients of
any 1 loop amplitude for any helicity configuration of the external particle. Since, implicitely, the method relies on
the calculation of tree amplitudes with various helicity configurations, we have also numerically implemented the 
BCFW recursion equation.

\section{The BCFW recursion equations}

In this section we briefly present the BCFW recursion and its numerical implementation. But first let us agree on 
some definitions and convention that we use throughout this work.

\subsection{Spinor definitions and conventions}
It is a well established fact how a massless four-vector can be written as matrix using the Pauli matrices 
$\sigma^{\mu}$:
\be
p_{\mu} \longrightarrow p_{A \dot B} = p_{\mu} \sigma^{\mu}_{ A \dot B}
\ee
Being a massless vector, the {\it matrix} $p_{A \dot B} $ has a determinant zero $ det(p)=0$. It is known from
linear algebra that a matrix with zero determinant can be written as a product of two columns which we call
spinors:
\be
p_{A \dot B} = \lambda_{A} \tilde{\lambda}_{\dot B}
\ee
We write these in a convienient bra-ket form as follows:
\be
\lambda_{ A}(p_i) \rightarrow |i \ket  \;\;\; , \;\;\;  \tilde{\lambda}_{\dot B}(p_i) \rightarrow \bra i |
\ee
For definite helicities we can write these spinors as solutions of the Weyl equation: 
\be
| i {\pm} \ket = u_{\pm}(p_i) \; \; \; \;  \bra  i {\pm} | = \bar{u}_{\pm}(p_i) 
\ee
With these definitions we can define inner products with spinors. There are two versions the `angle' product and the
`bracket'. For real momenta these are conjugate of each other. For complex momenta this may no longer hold.
\be
\bra i j \ket = \bra i- | j+ \ket = \bar{u}_- (p_i) u_+(p_j) \;\;\;\; \;\;\ 
[ij] =\bra i+ | j- \ket = \bar{u}_+ (p_i) u_-(p_j)
\ee
From the definitions of the spinors we can see that products like $\bra i+ | j+ \ket$ or $\bra i- | j- \ket$ vanish. We can
use these definitions to construct more complicated objects, with momenta sandwiched between spinors:
\be
\bra i + | k | j + \ket = \bra i +| \s k | j + \ket = [ik] \bra k j \ket  \;\;\;\; \;\;
\bra i - | k | j - \ket = \bra i -| \s k | j - \ket = \bra ik \ket  [ k j ]
\ee
with a generalization for $n$ momenta, 
\be
\bra i + | k_1 k_2 \cdots k_n | j + \ket = \bra i +| \s k_1 \cdots \s k_n | j + \ket = [ik_1] \bra k_1 k_2 \ket
[k_2 k_3] \cdots  \bra k_{n-1} k_n \ket
\ee
We will also use the identities:
\bea
&&\bra i + | k_1+k_2+\cdots+k_m+ \cdots+ k_n | j + \ket=\sum_{m=1}^n [im]\bra m j \ket \\
&&\bra i - | k_1+k_2+\cdots+k_m+ \cdots+ k_n | j - \ket=\sum_{m=1}^n \bra im \ket  [ m j ] 
\eea
Finally, as you may have already noticed, we will use the index of the momentum as the symbol for the spinor, when
there is no confusion.

\subsection{Analytic part}

For a LO amplitude, and concentrating on gluons only for simplicity, the color information can be factorized from the kinematical
part as follows:
\be
{\cal A}^{tree}_n ( \{ k_i,\lambda_i\} )= g^{n-2}   \sum_{ \sigma \in S_n/Z_n}
Tr \left(  T^{a_{\sigma(1)} } T^{a_{\sigma(2)} }\cdots T^{a_{\sigma(n)}} \right) 
 A_{n}^{tree} (\sigma(1^{\lambda_1}),\ldots, \sigma(n^{\lambda_n}) ) 
\ee
where $T^{a_i}$ are the $SU(N)$ matrices in the adjoint respresentation. The sum is over all { \it non-cyclic} permutations of
the external legs. The part of the factorization  $A_{n}^{tree} (\sigma(1^{\lambda_1}),\ldots, \sigma(n^{\lambda_n}) ) $ now
depends only on the external momenta, and the polarization vectors and is in a {\it color ordered} form. This has the nice
property that all poles that appear in it are made of adjacent momenta, $s_{12},s_{23}$ etc. An almost identical result holds for
quarks as well. 

Some years ago , Britto, Cachazo and Feng \cite{BCFW}, showed that the kinematical part of the amplitude, the so called { \it partial amplitude},  has a nice 
factorization property  that satisfies the recursion equation ( thereafter known as the BCFW recursion):
\be
A_{n}^{tree} (1,2, \ldots , n) = \sum_{h= \pm } \sum_{k=2}^{n-2} A_{k+1}^{tree} ( \hat 1,2, \ldots ,- \hat K_{1,k}^{-h} )
\frac{1}{K_{1,k}^{2}}  A_{n-k+1}^{tree} ( \hat K_{1,k}^{h}, \ldots , n-1 , \hat n )
\ee
The color ordered amplitude is split into two terms, consisting of lower point ordered amplitudes, joint by a propagator
$K_{1,k}=p_1 +p_2 + \ldots + p_k$. With an appropriate { \bf complex  } shift of the momenta the propagator can be put on shell.
The simplest way to do that is by shifting the first and the last leg (denoted by a hat in the recursion), by a fixed amount with a 
complex constant $z$:
\bea
&& 1 \rightarrow \hat 1 = 1 + z_k n \; \; \; , \; \; \;  \tilde 1  \rightarrow \tilde 1 \nonumber \\
&& n \rightarrow  n    \; \; \; , \; \; \; \; \; \; \; \; \;  \; \; \;  \; \; \;  \; \; \;  \;      \tilde n  \rightarrow \hat { \tilde n } = \tilde n  - z_k  \tilde 1
\eea
The $z$ is chosen in such a way so that the {\it shifted } propagator is massless:
\be
z_k = - \frac{K_{1,k}^2}{\bra n- | \s K_{1,k} | 1 - \ket}
\ee
\be
\hat K _{1,k}^2 = \left(  \s K _{1,k} + z_k n \tilde 1 \right)^2 =  K _{1,k}^2 +z_k \bra n- | \s K_{1,k} | 1 - \ket =0
\ee
The sums are over all possible distributions of the external momenta into the two groups, keeping the special, shifted momenta
in distinct groups, and over the helicities of the common propagator.
The recursion was proven by Britto et al \cite{BCFWProof1}  using the analytic properties of the amplitude and by
 Lazopoulos et al \cite{BCFWProof2} using standard Feynman diagrams analysis.

The advantage of the BCFW recursion is that  only a few terms are needed to compute a partial amplitude, compared to hundrends of
Feynman diagrams using standard field theory techniques. This signals major cancellations between graphs that are automatically 
taken care of using the recursion. Thus, we arrive at relatively compact expressions for the amplitudes. 

\subsection{Numerical implementation}
We have implemented the BCFW recursion, in a numerical { \tt FORTRAN90 } code, that computes the partial amplitudes
for QCD, including both quarks and gluons, for an arbitrary number of external legs. The only inputs in the code is the number of
legs $n$, the QCD process, and the helicity configuration. We have compared the results, both for MHV and NMHV helicity 
configurations with known results from the literature and with { \tt HELAC } \cite{HELAC}. We have also produced some new numerical results.
Some results are shown in the following tables for $8$ and $9$ point amplitudes with and without quarks. We have used 
the following momenta values for the $8$-point amplitude:
\bea
p_1 &=& (  50.0,0.0,0.0,50.0  ) \nonumber \\
p_2 &=& (  50.0,0.0,0.0,-50.0  ) \nonumber \\
p_3 &=& ( -11.1703767834197, -8.03205150805747 , 6.45958520425105 , -4.30548778241159) \nonumber \\
p_4 &=& ( -29.3374489041782 , -0.341091815726772 ,  5.38045260205076  ,28.8378274905652 ) \nonumber \\
p_5 &=& (  -8.93691541288138 , -7.79003576692047 ,  2.54732281477208  , -3.56299681791019 ) \nonumber \\
p_6 &=& ( -4.49061013802485 ,  -2.19138756179328  , 0.744707846345424 , 3.84822169181309 ) \nonumber \\
p_7 &=& ( -9.04125786194674 ,  4.56116168271723  , 7.76548604819915  , 0.798357229284189 ) \nonumber \\
p_8 &=& ( -37.0233908995491 ,13.7934049697808  ,  -22.8975545156185  ,-25.6159218113407  ) \nonumber 
\eea

\vspace*{0.3cm}

\begin{center}
\begin{tabular}{|l|c|}
\hline
 \bf Helicity Configuration & \bf Partial Amplitude \\ 
\hline
\hline
$\bf (1^- , 2^- ,3^ - , 4^-,5^+,6^+,7^+,8^+)$ & \small  $1.143995504515313 \; 10^{-4} + 2.452584716117081 \; 10^{-5} i $\\ 
\hline
 $\bf (1^- , 2^+ ,3^ + , 4^-,5^-,6^+,7^+,8^-)$ &  \small  $-3.892700175442802 \; 10^{-8} - 7.138278322124297 \; 10^{-9}  i $\\ 
\hline
$\bf (1^- , 2^- ,3^ - , 4^+,5^+,6^+,7^+,8^+)$ &  \small $2.973212107227238 \; 10^{-8}+ 1.495525686984027 \; 10^{-7} i $\\ 
\hline
$\bf (1^- , 2^- ,3^ + , 4^+,5^-,6^-,7^+,8^+)$ & \small $-8.620198212038414 \; 10^{-8} + 8.089172271501075 \; 10^{-6} i$\\ 
\hline
$\bf (1^+ , 2^+ ,3^ + , 4^-,5^+,6^-,7^+,8^-)$ &  \small $1.899386258601871 \; 10^{-6} -9.145607147903538 \; 10^{-6} i $\\ 
\hline
$\bf (1_q^- , 2_{\bar q}^+ ,3^ + , 4^+,5^+,6^-,7^-,8^-)$ & \small $-1.812332579250756\; 10^{-7}+ 1.534899358766671 \; 10^{-7} i $ \\ 
\hline
$\bf (1_q^- , 2_{\bar q}^+ ,3^ - , 4^+,5^-,6^+,7^-,8^+)$  & \small $ -5.724656620620708 \; 10^{-9} -8.875798503634412\; 10^{-10} i $ \\
\hline
\end{tabular}
\end{center}
\begin{center}
{ \footnotesize { \bf Table 1. } Results for 8-point partial amplitudes. Quarks and antiquarks are denoted as subscripts $q$ and $\bar q $ 
in the relevant legs }
\end{center}
The following momenta configuration was used for the $9$-point amplitudes:
\bea
p_1 &=& (  50.0,0.0,0.0,50.0  ) \nonumber \\
p_2 &=& (  50.0,0.0,0.0,-50.0  ) \nonumber \\
p_3 &=& (-8.62171654444322 , -6.90534943082250 ,  4.59215344257906  ,-2.35844698282037 ) \nonumber \\
p_4 &=& ( -27.8413520112470 , -0.643296108254438 , 0.840299169363624  , 27.8212319863308 ) \nonumber \\
p_5 &=& ( -7.13369828021707 , -6.67842455614503 ,  1.44431114936003  , -2.04994192719828 ) \nonumber \\
p_6 &=& ( -4.22769845708132,  -1.90930333047348 ,  7.776353148874293 \; 10^{-2}, 3.77119979176668 ) \nonumber \\
p_7 &=& ( -7.10934176699029 , 3.77076042627538  ,  5.74622866522395  , 1.81795552598701 ) \nonumber \\
p_8 &=& ( -30.9459371413082 , 11.4021700192282  , -22.0566594617972  , -18.4701195933519) \nonumber \\
p_9 &=& ( -14.1202557987129 , 0.963442980191897 ,  9.35590350378182  , -10.5318788007139 ) \nonumber 
\eea
\vspace*{0.3cm}
\begin{center}
\begin{tabular}{|l|c|}
\hline
 \bf Helicity Configuration & \bf Partial Amplitude \\ 
\hline
\hline
$\bf (1^- , 2^- ,3^ - , 4^-,5^+,6^+,7^+,8^+,9^+)$ &  \small $-1.195041826437500 \; 10^{-6}-4.267905334966595 \; 10^{-6} i $\\ 
\hline
 $\bf (1^- , 2^+ ,3^ + , 4^-,5^-,6^+,7^+,8^-,9^-)$ &  \small $9.960596678335883 \; 10^{-9}+ 2.632108576287432 \; 10^{-10} i $\\ 
\hline
$\bf (1^- , 2^- ,3^ - , 4^+,5^+,6^+,7^+,8^+,9^+)$ &  \small  $-2.628610959245734 \; 10^{-10}-3.014249899217379 \; 10^{-9} i $\\ 
\hline
$\bf (1^- , 2^- ,3^ + , 4^+,5^-,6^-,7^+,8^+,9^-)$ & \small $-8.470520978313771 \; 10^{-7}-1.528306470088502 \; 10^{-6} i $ \\ 
\hline
$\bf (1^+ , 2^+ ,3^ + , 4^-,5^+,6^-,7^+,8^-,9^+)$  & \small $-1.935502160225642 \; 10^{-7} + 7.283711249910175 \; 10^{-8}i $ \\ 
\hline
$\bf (1_q^- , 2_{\bar q}^+ ,3^ + , 4^+,5^+,6^-,7^-,8^-,9^-)$ & \small  $ -4.751617821378328 \; 10^{-8} -5.102361545628036 \; 10^{-8} i $ \\ 
\hline
$\bf (1_q^- , 2_{\bar q}^+ ,3^ - , 4^+,5^-,6^+,7^-,8^+,9^-)$ & \small $-4.359890532350093 \; 10^{-9}+ 1.991698153552900 \; 10^{-9} i $\\
\hline
\end{tabular}
\end{center}
\begin{center}
{ \footnotesize { \bf Table 2. } Results for 9-point partial amplitudes. Notations are the same as before}
\end{center}

\section{The next step: NLO}

The color factorization structure for the LO amplitudes, persists for NLO amplitudes only slightly more
involved:
\bea
{\cal A}^{1-loop}_n ( \{ k_i,\lambda_i\} )&=& g^n   \sum_{ \sigma \in S_n/Z_n}
 G_{n;1}(\sigma) A_{n;1}
(\sigma(1^{\lambda_1}),\ldots, \sigma(n^{\lambda_n}) )  \nl
 &+& g^n \sum_{c=2}^{n/2+1} \sum_{ \sigma \in S_n/S_{n;c}}  G_{n;c}(\sigma) 
  A_{n;c} (\sigma(1^{\lambda_1}),\ldots, \sigma(n^{\lambda_n}) ) 
\eea
where $G_{n;1}(\sigma)$ is the leading color structure:
\be
G_{n;1}(\sigma) = N_c \; Tr \left(  T^{a_{\sigma(1)} } T^{a_{\sigma(2)} }\cdots T^{a_{\sigma(n)}} \right) 
\ee
and $G_{n;c}(\sigma)$ is the subleading color structure:
\be
G_{n;c}(\sigma)=Tr \left(  T^{a_{\sigma(1)} } \cdots T^{a_{\sigma(c-1)}} \right) 
Tr \left(  T^{a_{\sigma(c)} } \cdots T^{a_{\sigma(n)}} \right) 
\ee
The colored ordered amplitudes of the subleading terms,  $A_{n;c} (\sigma )  $,  can be written as linear combinations of permumations of the leading color amplitudes
$A_{n;1} (\sigma ) $, so the latter are called  {\it primitive amplitudes}. For the rest of this paper we focus
on these primitive amplitudes.

By restricting ourselves to cyclicaly ordered primitive amplitudes we reduce the labour of computing 
hundrends or thousands of Feynman diagrams to a reduced set of diagrams. Now each of those diagrams
is an integral over the loop momentum. The integrands of those integrals are made of tensors or vectors of
the external momenta and the loop momenta. It would an enormous simplification if
we could further reduce these to a basic set of integrals, so that every diagram could be written as a combination
in this basis. It turns out that such a reduction is possible, either using standard Passarino-Veltman techniques 
\cite{Passarino}, or
the more recent Ossola-Papadopoulos-Pittau \cite{OPP} reduction method. Thus all loop integrals can be brought down to a basis consisting of up to 4 propagators:

\be
{\cal I} = \{ I_2,I_3^{1m},I_3^{2m},I_3^{3m},I_4^{1m},I_4^{2m \; e}, I_4^{2m \; h}
    , I_4^{3m},I_4^{4m} \}
\ee
As a result any 1 loop amplitude can be written as a linear combination in this basis, with {\it algebraic} coefficients.

\be
A_n^{1-loop} \sim \sum_{j} a_j {\cal I}_j + {\cal R }_n
\ee
where $ {\cal R }_n$ are the rational terms. More explicitly we can write:

\be
A_n^{1-loop} \sim \sum_i b_i B(K_i^2) + \sum_{ij} c_{ij} C(K_i^2,K_j^2) 
                       + \sum_{ijk} d_{ijk} D(K_i^2,K_j^2,K_k^2) +{\cal R }_n
\ee
where $B,C$ and $D$ are the bubble, triangle and box integrals respectively. Analytic expressions and singularity structure for these
integrals can be found for example in \cite{vanHameren:2005ed} 

\section{Generalized Unitarity}

Unitarity of the scattering matrix in field theory implies  the conservation of probability. On the amplitude level
unitarity says that cutting a loop gives the discontinuity in the scattering amplitude. By cutting we mean putting a
propagator on shell, which amounts in the replacement:
\be
\frac{i}{p^2 +i \epsilon} \to 2 \pi \delta^{(+)}(p^2)
\ee
At the diagrammatic level this the well known Cutkosky rule. In general, at the amplitude level, cutting an amplitude
in a given channel isolates those integrals that have a discontinuity across that channel. In other words cutting gives
a linear combination of master integrals with algebraic coefficients. If we wish to isolate one single integral we must go a bit
further and insist {\it more} propagators go on shell. This goes under the name of {\it generalized unitarity} 
\cite{GenUni}. For the box
integrals in particular this is quite easy to do. Cutting four propagators, isolates a single box integral. The cut breaks the loop
integral in four tree level amplitudes (one at each corner of the box) and the coefficient of the box integral is simply the
product of those tree amplitudes:
\vspace{0.5cm} 
\begin{center}
\framebox[1.3\width]{
\includegraphics[width=0.38\textwidth]{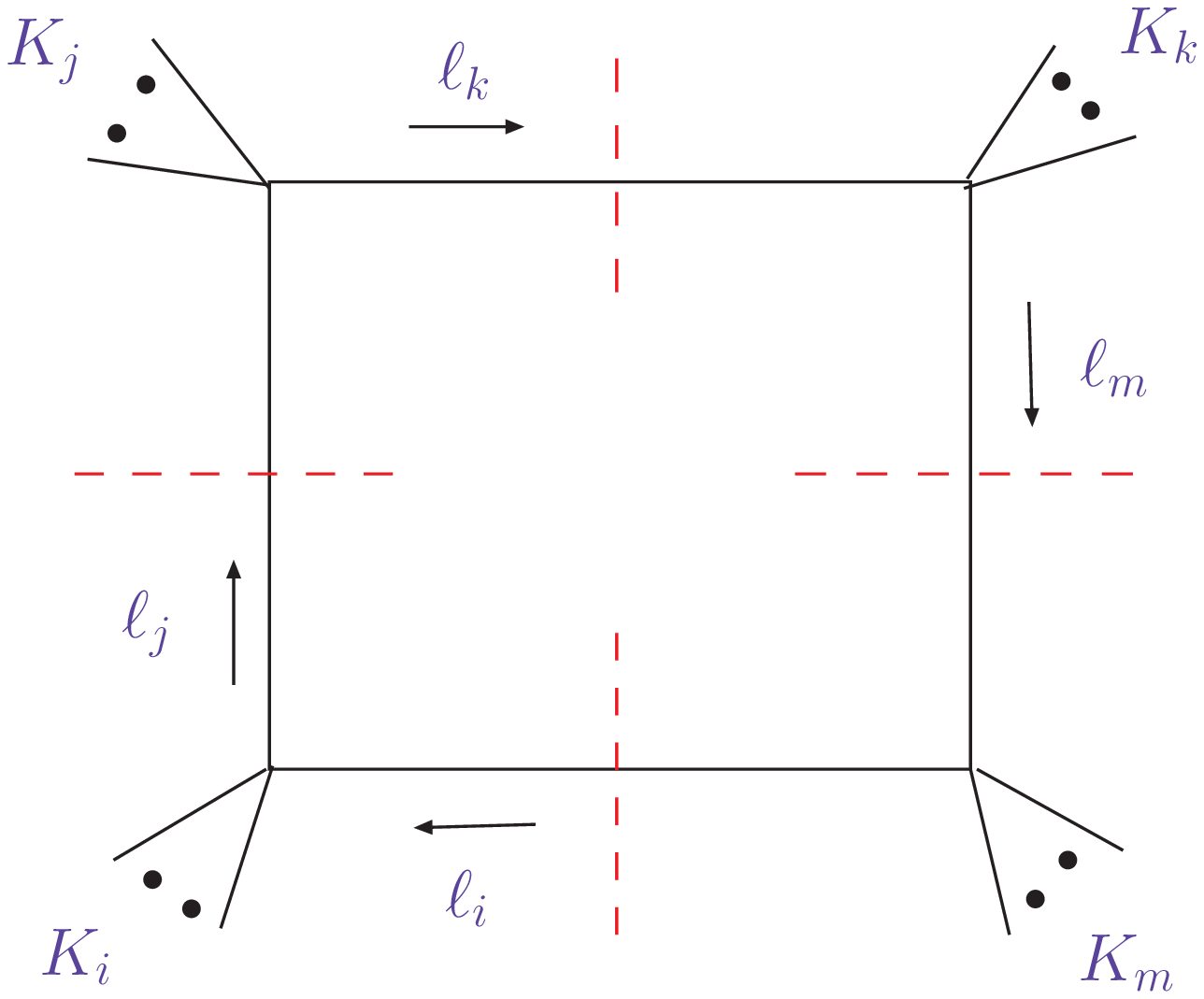} } \\
\vspace{0.2cm} 
 {\small { \bf Fig.1 } Cutting 4 propagators, gives the Box coefficient, as a product of the tree level}
                                    
\end{center}   
\vspace{0.5cm} 
\bea
d_{ijk}= &\frac{1}{2} & \sum_{a=1}^2 \sum_{h=\pm} A_1^{tree} ( \ell_i^{h_i}, \left\lbrace i \right\rbrace , -\ell_j^{-h_j})
A_2^{tree} ( \ell_j^{h_j}, \left\lbrace j \right\rbrace , -\ell_k^{-h_k })  \nonumber \\
&& \times A_3^{tree} ( \ell_k^{h_k}, \left\lbrace k \right\rbrace , -\ell_m^{-h_m})
A_4^{tree} ( \ell_m^{h_m}, \left\lbrace m \right\rbrace , -\ell_i^{-h_i})
\label{Coeff}
\eea
where $\left\lbrace i \right\rbrace, \left\lbrace j \right\rbrace$ etc., denotes the collection of momenta in that vertex of the box.
One also averages over the two solutions of the quadratic equation one obtains by putting four propagators on shell. Finally there is a sum
over the helicities of the propagators.

\paragraph{Simple example $A(1^+,2^+,3^-,4^-)$} 

We have to satisfy the on-shell conditions: $ l_1^2=0, l_2^2=0,l_3^2=0,l_4^2=0$. Choosing $l_2$ to be :
$l_2^{\mu}= \xi  \bra 1 - | \gamma^{\mu} | 2- \ket $ , where $\xi$ is a normalization, satisfies the first
three. The last will determine  $\xi$. The product of the trees in the corners of the box is:
\be
\frac{1}{2}
\frac{[1l_2]^3}{[l_2 l_1] [l_1 1 ] } \frac{\bra l_4 4 \ket^3}{\bra 4 l_1 \ket \bra l_1 l_4 \ket }
\frac{\bra l_3 l_2 \ket^3}{\bra  l_2 2 \ket \bra 2 l_3 \ket }
\frac{[l_4 l_3]^3}{[l_3 3] [3 l_4]}
\ee

We can combine the numerator into one factor:
\be
[1 l_2] \bra l_2 l_3 \ket [ l_3 l_4 ] \bra l_4 4 \ket = \bra 1 + | l_2 l_3 l_4 | 4+ \ket
\ee
Using the identities $\s l_3 \s l_4 =\s l_3 ( \s l_3 +\s 3)= \s l_3 \s 3$ and 
$\s l_2 \s l_3 \s l_4 =\s l_2 \s l_3 \s 3 =\s l_2 (\s l_2 +\s 2 ) \s 3 =\s l_2 \s 2 \s 3 $ this can be simplified:
\be
\bra 1 + | l_2 l_3 l_4 | 4+ \ket = \bra 1 + | l_2 2 3 | 4+ \ket = \bra 1 + | l_2 | 2+ \ket [23] \bra 3 4 \ket 
= s_{12} [23] \bra 3 4 \ket
\ee
Combining appropriately factors in the denominator, we can simplify the expression:
\be
[3 l_4] \bra l_4 l_1 \ket [ l_1 l_2 ] \bra l_2 2 \ket = \bra 3 + | l_4 l_1 l_2 | 2+ \ket 
=[34] \bra 4 1 \ket  \bra 1 + | l_2 | 2+ \ket =  [34] \bra 4 1 \ket s_{12}
\ee
\be
\bra 4 l_1 \ket [l_1 1 ] = \bra 4 - | l_1 | 1- \ket = \bra 4 1 \ket [21] 
\ee
\be
\bra 2 l_3 \ket [l_3 3 ] = \bra 2 - | l_3 | 3- \ket = \bra 2 1 \ket [23] 
\ee
Putting together numerator and denominator and using conservation of momenta we finally get:
\be
d_{1:2:3:4} = \frac{1}{2} s_{12} s_{23} A_4^{tree} (1^+,2^+,3^-,4^-)
\ee

\section{Numerical Implementation}

We have implemented generalized unitarity in a numerical {\tt FORTRAN 95} code. For the
time being it has been implemented for box coefficients only and for purely gluonic
amplitudes (no quarks in the loop). The evaluation of the box coefficient goes through
the following steps:
\begin{itemize}
 \item {\bf Lists all possible groupings for the external momenta:} The external momenta
      are grouped in 4 groups (the corners of the box), in all possible, distinct ways.
      Sums of these momenta make up the channel that the cut is computed.
 \item {\bf Solutions of the onshell loop momenta condition:} The system of equations:
      $\ell_1^2=0, \ell_2^2=0, \ell_3^2=0, \ell_4^2=0$ for the cut loop momenta is solved
      and the two solutions, that we have to average over are evaluated
 \item {\bf The tree amplitudes are computed:} For each corner of the box we compute the
     tree amplitude defined by the external momenta in that corner, the solution for the
     loop momentum that we just obtained and the particular helicity configuration of the
     grouping of the external momenta. The computation of the tree amplitudes is done 
     automatically using the numerical code for the BCFW recursion that we discussed
     in a previous section.
 \item Finally the coefficient for that particular cut is computed, using 
       Eq. (\ref{Coeff}).
\end{itemize}

We have produced some results for two 1-loop processes, namely $gg \to 8g$ and $gg \to 9g$
for various choices of helicity configurations and for various channels (particular 
coefficients of a master integral). We have used the same momenta configuration as in the
BCFW recursion section. We have compared our results with the ones in the literature and we have also produced some new numerical results in the case of 1-loop $9$ gluon amplitudes. The results are shown in the following tables. 

\begin{center}
\begin{tabular}{|c|c|c|}
\hline 
  & \scriptsize { $\mathbf{-+-+-+-+}$}& \scriptsize{$\mathbf{----++++}$ }  \\ 
\hline 
\hline 
\scriptsize{$\mathbf{(1:2:3:45678)}$} & \scriptsize{$-0.452833202541873+1.83433531772824  i$} &
          \scriptsize{ $564.146943978796 + 341.450920347787 i $ }   \\ 
\scriptsize{$\mathbf{(1:23:4:5678)} $} & \scriptsize{$-1.511201093924664 10^{-5} -2.692526296296084 10^{-5} i $}
   &     \scriptsize{ $85.7803228327941 -4.79606435167831 i $}    \\ 
\scriptsize{$\mathbf{ (1:2:345:678)}$} & \scriptsize{$-0.596872851993693-1.13690332085509 i$}& 
        \scriptsize{ $366.912336610440 +98.0271230853793 i $}  \\
\scriptsize{$\mathbf{(1:2:34:5678)} $} & \scriptsize{$0.343744213463153-0.116177716850307 i $} & 
        \scriptsize{$0.0 + 0.0 i $} \\
\scriptsize{$\mathbf{(12:34:56:78)} $} &  \scriptsize{$-6.089365062436733 10^{-3} -0.185567043177724 i $ } &  
        \scriptsize{$0.0 + 0.0 i $}  \\
\hline
\end{tabular}
\end{center}

\begin{flushleft}
\begin{tabular}{|c|c|c|}
\hline 
  & \scriptsize{$\mathbf{-+-+-+++}$}  & \scriptsize{$\mathbf{-+-+++-+}$}\\ 
\hline 
\hline 
\scriptsize{$\mathbf{(1:2:3:45678)}$} & \scriptsize{$ -1.810002621799826 \; 10^{-2}-7.007553695597921 \; 10^{-5} i$} &
          \scriptsize{ $1.282527393614916 \; 10^{-3}-1.544463085586650 \; 10^{-2} i $ }   \\ 
\scriptsize{$\mathbf{(1:23:4:5678)} $} & \scriptsize{$-9.061239878511741 \; 10^{-7}-6.182682488354774 \; 10^{-7} i $}
   &     \scriptsize{ $-3.593566469301429 \; 10^{-6} -2.807114551959228 \; 10^{-6} i $}    \\ 
\scriptsize{$\mathbf{ (1:2:345:678)}$} & \scriptsize{$ -6.198557976336275 \; 10^{-7} -3.145711006839414 \; 10^{-7}i$}& 
        \scriptsize{ $ 1.201810688581596 \; 10^{-3}+ 1.747770080527506 \; 10^{-4}i $}  \\
\scriptsize{$\mathbf{(1:2:34:5678)} $} & \scriptsize{$2.057234610074154 \; 10^{-3}+5.080380495579885 \; 10^{-5} i $} & 
        \scriptsize{$0.0 + 0.0 i $} \\
\scriptsize{$\mathbf{(12:34:56:78)} $} & \scriptsize{$0.0 + 0.0 i $} &  
        \scriptsize{$0.0 + 0.0 i $}  \\
\hline
\end{tabular}
\end{flushleft}
\begin{center}
{ \footnotesize { \bf Table 3. } Results for 1-loop, 8-point primitive amplitudes.}
\end{center}

The code provides {\bf all} the coefficients for a particular process at once. Typical times
for the two processes shown in the tables are of the order of $1/10$th of a second on a 64-bit Intel Core Duo at 2GHz.

\newpage

\begin{center}
\begin{tabular}{|c|c|c|}
\hline 
  & \scriptsize { $\mathbf{-+-+-+-+-}$}& \scriptsize{$\mathbf{----+++++}$ }  \\ 
\hline 
\hline 
\scriptsize{$\mathbf{(1:2:3:456789)}$} & \scriptsize{$  -0.291597431159727 +0.263428066914340 i$} &
          \scriptsize{ $-9.42379749599236 +8.63580103044842 i $ }   \\ 
\scriptsize{$\mathbf{(1:23:4:56789)} $} & \scriptsize{$6.917959935625083 \; 10^{-7}-2.862781163146907 \; 10^{-7} i $}
   &     \scriptsize{ $-4.424679500634447 \; 10^{-2}-1.06642829663490 i $}    \\ 
\scriptsize{$\mathbf{ (1:2:345:6789)}$} & \scriptsize{$1.83770214755652 +1.25751379616759 i$}& 
        \scriptsize{ $-3.16966894074953 -10.5119953247915 i $}  \\
\scriptsize{$\mathbf{(1:2:34:56789)} $} & \scriptsize{$-0.801669928958509 -0.704349268560792  i $} & 
        \scriptsize{$0.0 + 0.0 i $} \\
\scriptsize{$\mathbf{(12:34:56:789)} $} &  \scriptsize{$4.147713723224443 \;10^{-2}+3.388946250063977 \; 10^{-2} i $ } &  
        \scriptsize{$0.0 + 0.0 i $}  \\
\hline
\end{tabular}
\end{center}

\begin{flushleft}
\begin{tabular}{|c|c|c|}
\hline 
  & \scriptsize{$\mathbf{-+-+-+++-}$}  & \scriptsize{$\mathbf{-+-+++-+-}$}\\ 
\hline 
\hline 
\scriptsize{$\mathbf{(1:2:3:456789)}$} & \scriptsize{$-2.382052305295742 \; 10^{-3}-1.320540695786225 \; 10^{-2}  i$} &
          \scriptsize{ $7.517076253761092 \; 10^{-3}-2.651249643255122 \; 10^{-3} i $ }   \\ 
\scriptsize{$\mathbf{(1:23:4:56789)} $} & \scriptsize{$2.330322361904153 \; 10^{-8}-2.184457500759925 \; 10^{-8} i $}
   &     \scriptsize{ $7.985298353983870 \; 10^{-8}-1.068015483724008 \; 10^{-7} i $}    \\ 
\scriptsize{$\mathbf{ (1:2:345:6789)}$} & \scriptsize{$ 1.731650767307649 \; 10^{-2} -1.381732810170124 \; 10^{-2} i$}& 
        \scriptsize{ $1.377043202412121 \; 10^{-4}+1.023698942546792 \; 10^{-4} i $}  \\
\scriptsize{$\mathbf{(1:2:34:56789)} $} & \scriptsize{$-9.470193967875743^10^{-3}+8.652362025481619 \; 10^{-3} i $} & 
        \scriptsize{$ -5.650383858159608 \; 10^{-4} -1.479340069213934 \; 10^{-4} i $} \\
\scriptsize{$\mathbf{(12:34:56:789)} $} & \scriptsize{$ -8.488224296845019 \; 10^{-5}-1.761227642104737 \;10^{-4} i $} &  
        \scriptsize{$3.287516684614978 \; 10^{-8} +1.134952815474561 \; 10^{-7} i $}  \\
\hline
\end{tabular}
\end{flushleft}
\begin{center}
{ \footnotesize { \bf Table 4. } Results for 1-loop, 9-point primitive amplitudes}
\end{center}

Triangle coefficients can be obtained in much the same way. Cutting 3 propagators does not immediately isolate a single integral
but it gives a triangle integral plus a sum of box integrals \cite{Forde:2007mi}. This is because the particular box integrals share the same cuts
with the triangles. The boxes
have to be subtracted in a suitable way. This can naturally be implemented in our algorithm since box coefficients are computed already.
For bubble integrals, cutting 2 propagators and subtracting the triangle and box contributions will isolate a single coefficient. Work is in
progress to implement just this procedure.

\section{Summary and Outlook}
We presented numerical implementations for the BCFW recursion equations and 1-loop coefficients of box integrals using generalized unitarity. In the case of the BCFW recursion equations we were able to provide, in a fast algorithmic way, multi-particle amplitudes both old and new, namely in the case of 9 parton Leading Order amplitude. For the implementation of generalized unitarity we are able to compute all box coefficients at
once for an arbitrary number of particles and helicity configuration. Some new coefficients for the 1 loop 9 gluon amplitude were presented.
 Work is in progress to include quarks in the picture (easy step since this is already done in the BCFW recursion code)
and also complete the implementation by computing triangle and bubble coefficients. 

\section{Acknowledgements}

I wish to thank Costas Papadopoulos for discussions and  useful comments. \\
Author P.D. is co-funded by the European
Social Fund (75\%) and National Resources (25\%)-EPEAEK B! - PYTHAGORAS.

\end{document}